\documentclass[preprint,amsmath,amssymb]{revtex4}
\usepackage{graphicx}
\usepackage{bm}
\makeatletter 
\renewcommand{\fnum@figure}{\textbf{Figure~\thefigure}}
\makeatother

\begin{document}

\title{Suppression of the antiferromagnetic pseudogap in the electron-doped high-temperature superconductor by \\ ``protect annealing" \vspace{1cm}}

\author{M. Horio$^{1*}$, T. Adachi$^2$, Y. Mori$^3$, A. Takahashi$^3$, T. Yoshida$^1$, H. Suzuki$^1$, L. C. C. Ambolode II$^1$, K. Okazaki$^1$, K. Ono$^4$, H. Kumigashira$^4$,  H. Anzai$^{5\dag}$, M. Arita$^5$, H. Namatame$^5$, M. Taniguchi$^{5,6}$, D. Ootsuki$^1$, K. Sawada$^7$, M. Takahashi$^7$, T. Mizokawa$^7$, Y. Koike$^3$, and A. Fujimori$^{1}$} \noaffiliation

\begingroup
\let\clearpage\relax
\let\vfil\relax
\maketitle
\endgroup

\noindent
{\it \footnotesize $^1$Department of Physics, University of Tokyo, Bunkyo-ku, Tokyo 113-0033, Japan}  \\
{\it \footnotesize $^2$Department of Engineering and Applied Sciences, Sophia University, Tokyo 102-8554, Japan} \\
{\it \footnotesize $^3$Department of Applied Physics, Tohoku University, Sendai 980-8579, Japan} \\
{\it \footnotesize $^4$KEK, Photon Factory, Tsukuba 305-0801, Japan} \\
{\it \footnotesize $^5$Hiroshima Synchrotron Radiation Center, Hiroshima University, Higashi-Hiroshima 739-0046, Japan} \\
{\it \footnotesize $^6$Graduate School of Science, Hiroshima University, Higashi-Hiroshima 739-8526, Japan} \\
{\it \footnotesize $^7$Graduate School of Frontier Sciences, University of Tokyo, Kashiwa 277-0882, Japan} \\

\vfill

\noindent
$^*$e-mail: horio@wyvern.phys.s.u-tokyo.ac.jp \\
{\footnotesize $^{\dag}$Present address: Graduate School of Engineering, Osaka Prefecture University, Sakai 599-8531, Japan \\}

\newpage
{\bf In the hole-doped cuprates, a small amount of carriers suppresses antiferromagnetism and induces superconductivity. In the electron-doped cuprates, on the other hand, superconductivity appears only in a narrow range of high electron concentration ($\sim$ doped Ce content) after reduction annealing, and strong antiferromagnetic (AFM) correlation persists in the superconducting phase. Recently, Pr$_{\bm{1.3-x}}$La$_{\bm{0.7}}$Ce$_{\bm{x}}$CuO$_{\bm{4}}$ (PLCCO) bulk single crystals annealed by a ``protect annealing" method showed a high $\bm{T_c}$ of $\sim$ 27 K for small Ce content down to $\bm{x \sim 0.05}$. By angle-resolved photoemission spectroscopy (ARPES) measurements of PLCCO crystals, we observed a sharp quasi-particle peak on the entire Fermi surface without signature of an AFM pseudogap unlike all the previous work, indicating a dramatic reduction of AFM correlation length and/or of magnetic moments. The superconducting state was found to extend over a wide electron concentration range. The present ARPES results fundamentally change the long-standing picture on the electronic structure in the electron-doped regime.}

\begin{figure}
	\begin{center}
		\includegraphics[width=140mm]{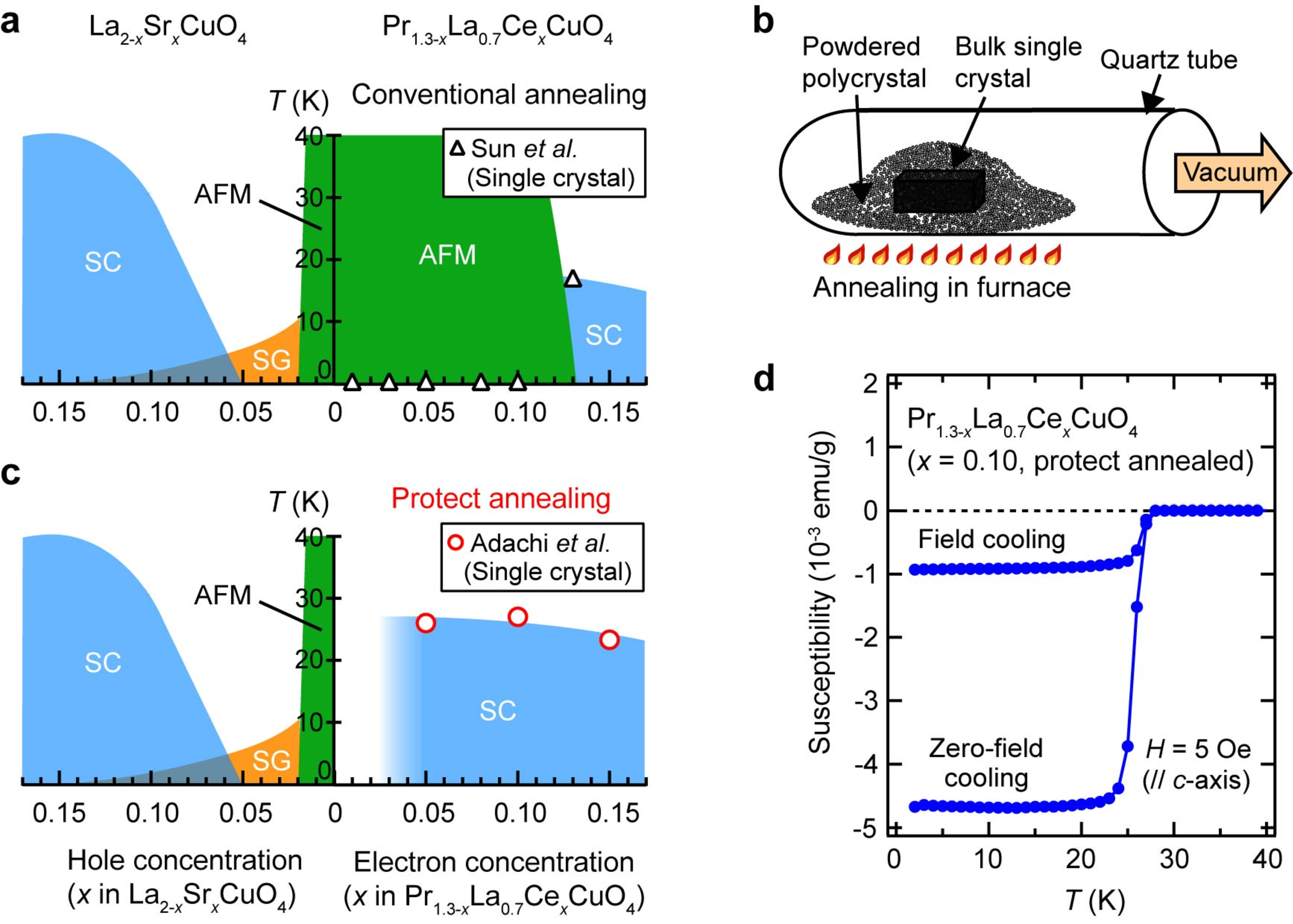}	
	\end{center}
	\caption{{\bf Superconducting properties of Pr$_{1.3-x}$La$_{0.7}$Ce$_x$CuO$_4$ (PLCCO) samples.} {\bf a,} $T_c$'s determined from the resistivity of PLCCO single crystals annealed by the conventional method reported by Sun {\it et al.} (open triangles) \cite{Sun}. The antiferromagnetic (AFM) and superconducting phases are denoted by AFM and SC, respectively. A typical phase diagram for a hole-doped cuprate La$_{2-x}$Sr$_x$CuO$_4$ is also shown on the left-hand side. {\bf b,} Schematic description of the protect annealing method. {\bf c,} The same plot as {\bf a} for protect-annealed single crystals reported by Adachi {\it et al.} (T.A., A.T., M. Ohgi, and Y.K. unpublished., open circles). $T_c$ was determined from magnetic susceptibility measurements. {\bf d,} Magnetic susceptibility of a protect-annealed PLCCO single crystal which shows the $T_c$ of 27.0 K.}
	\label{PLCCO_phase}
\end{figure}

Since the discovery of the cuprate high-temperature superconductors, one of the central issues has been the relationship between antiferromagnetic (AFM) order or AFM spin fluctuations and superconductivity. Starting from the AFM parent insulator, a small amount ($\sim$2\%) of hole doping destroys the AFM ordering and superconductivity emerges. However, for the electron-doped high-temperature superconductors (e-HTSCs), the antiferromagnetism persists up to the optimum doping ($\sim$ 15\%), as depicted in Fig.~\ref{PLCCO_phase}a. In the underdoped region of e-HTSCs, a large pseudogap opens due to AFM order or AFM correlation as observed by optical measurements \cite{Onose.optical,Wang} and scanning tunneling spectroscopy (STS) \cite{Zimmers}. Angle-resolved photoemission (ARPES) studies have shown that the ``pseudogap" opens around the ``hot spots", namely, crossing points of the Fermi surface (FS) with the AFM Brillouin zone boundary in superconducting samples \cite{armitage1,Matsui.anisotropy,Matsui.SCgap}. A neutron scattering study \cite{Motoyama} has revealed that the AFM correlation length is of order $\sim 10$ lattice spacing in the superconducting phase, and the ``pseudogap" in the ARPES spectra of the superconducting phase has been reproduced by assuming a similar AFM correlation length \cite{Park.short.range,Park_phase_fluctuation}.

Since the discovery of the e-HTSCs, it has been well known that annealing in a reducing atmosphere is necessary to realize superconductivity. As-grown samples are non-superconducting and AFM. By annealing the AFM phase shrinks, and superconductivity appears \cite{Tokura.NCCO}. It has been believed that a small amount of residual oxygen atoms at the apical oxygen site causes the AFM ordering, and are removed by reduction annealing \cite{Radaelli.apical}. Previous ARPES studies have revealed that reduction annealing decreases the intensity of the AFM folded bands and increase the spectral intensity at Fermi level ($E\mathrm{_F}$) \cite{Richard, Dongjoon}, but the AFM pseudogap has been seen in all the e-HTSCs from the underdoped to overdoped regions studied so far \cite{Matsui.doping}. Therefore, the AFM pseudogap has been regarded as a hallmark of the e-HTSCs and the relationship between antiferromagnetism and superconductivity has been considered as a more essential ingredient of the e-HTSCs than the hole-doped ones. \\

In a previous study, Brinkmann {\it et al.} \cite{Brinkmann} annealed thin single crystals of Pr$_{2-x}$Ce$_x$CuO$_4$ (PCCO) sandwiched by PCCO polycrystals of the same compositions and realized superconductivity with Ce concentration as low as 4\%. Recently, in thin films \cite{Tsukada.Nondoped, Matsumoto.NCCO} and powdered samples \cite{Asai.powder, Takamatsu.powder} of e-HTSCs, superconductivity was found even without Ce doping. Inspired by those studies, Adachi {\it et al.} \cite{Adachi} further improved the reduction annealing method of Brinkmann {\it et al.} by using powders instead of polycrystals as schematically shown in Fig.~\ref{PLCCO_phase}b, and successfully synthesized bulk superconducting single crystals of Pr$_{1.3-x}$La$_{0.7}$Ce$_x$CuO$_4$  (PLCCO) with $x = 0.10$. They call this new reduction annealing method ``protect annealing" method. Although PLCCO samples with such a low Ce concentration did not show superconductivity in previous studies \cite{Sun} (Fig.~\ref{PLCCO_phase}a), the protect-annealed samples showed a $T_c$ as high as 27.0 K (even higher than those prepared by conventional annealing) as shown in Figs.~\ref{PLCCO_phase}c and d (T.A., A.T., M. Ohgi, and Y.K. unpublished.). In order to study the effect of protect annealing on the electronic structure, we have performed ARPES measurements on single crystals of PLCCO with $x = 0.10$ with varying annealing conditions (see Methods). Presented results for the ``annealed sample" are those for ``annealed sample 1" with a $T_c$ of 27.0 K out of three protect-annealed superconducting samples unless otherwise stated.

\begin{figure}
	\begin{center}
		\includegraphics[width=140mm]{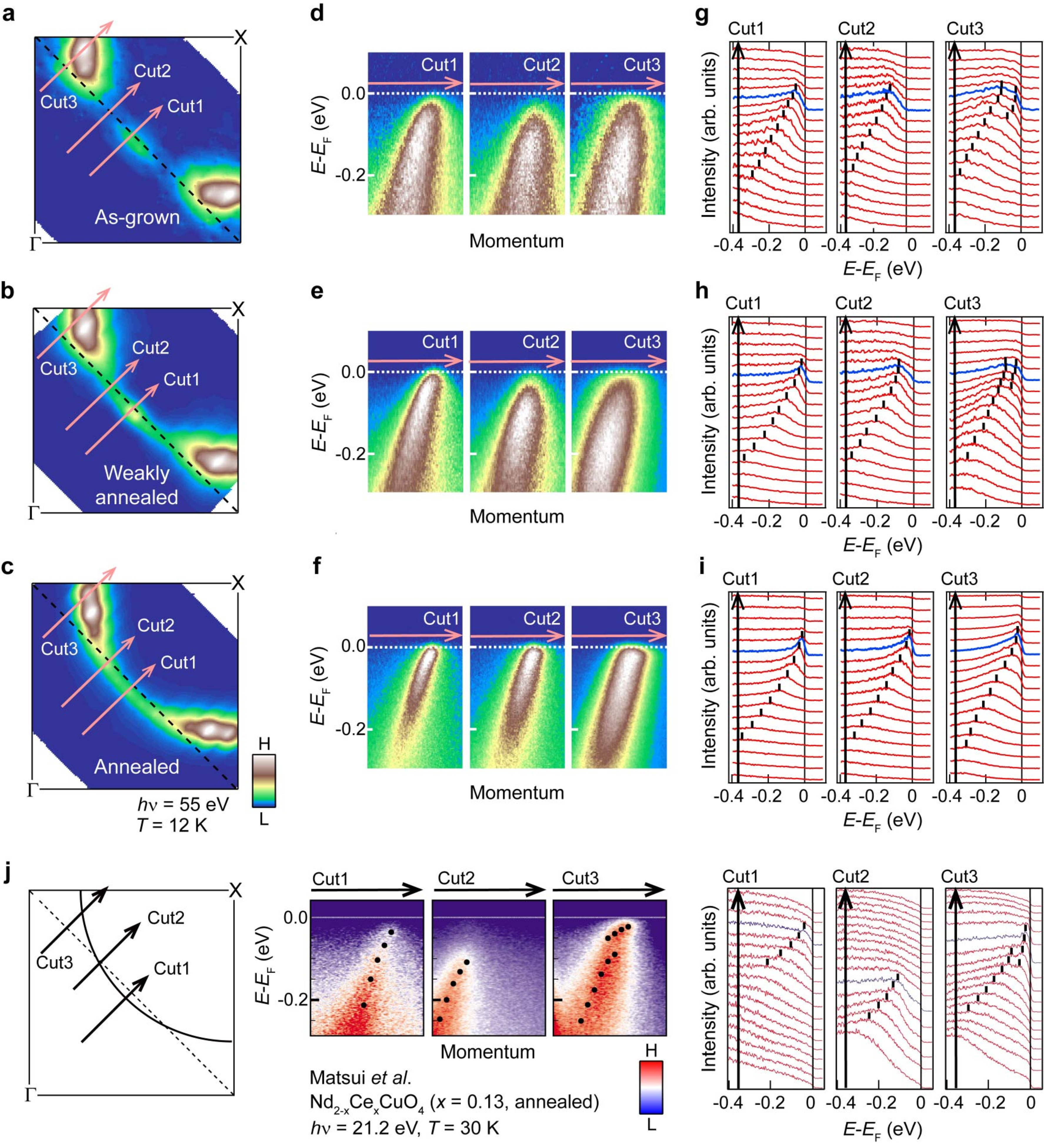}
	\end{center}
	\caption{{\bf ARPES spectra of PLCCO with and without protect annealing.} {\bf a-c,} Fermi surface mappings of as-grown, weakly annealed, and annealed samples, respectively. The intensity is integrated over $\pm 10\mathrm{meV}$ of $E\mathrm{_F}$. The suppressed intensities at the ``hot spots", the crossing points of the Fermi surface and the AFM Brillouin zone boundary, are fully recovered in the annealed sample. {\bf d-f,} Intensity plot in energy-momentum space for each sample along each cut indicated in {\bf a}. {\bf g-i,} EDCs plotted for each cut. Blue EDCs are taken at $k\mathrm{_F}$'s. Peak positions are marked by vertical bars. The AFM pseudogap that is observed at the ``hot spots" (cut 2) of the as-grown and weakly annealed samples is suppressed in the annealed sample. {\bf j}, Corresponding plots to {\bf d-f} and {\bf g-i} taken from ref.~\cite{Matsui.anisotropy} for Nd$_{2-x}$Ce$_x$CuO$_4$ ($x = 0.13$) annealed by the conventional method. Cut positions are indicated in the left panel.}
	\label{fig2}
\end{figure}

In Figs.~\ref{fig2}a-c, FS mappings of as-grown, weakly annealed (non-superconducting), and annealed ($T_c = 27.0$ K) samples are shown. In the as-grown sample, the intensities are strongly suppressed around the ``hot spots" due to the AFM order. The intensity partially recovers by the weak annealing, but the FS is still disconnected between the nodal and anti-nodal regions by the presence of the ``hot spots". This means that the weak annealing was not enough for the removal of apical oxygen and the influence of AFM correlation still persists. However, in the sufficiently annealed sample, the suppressed intensities at the ``hot spots" were fully recovered, and the entire FS became a continuous circular one. This very simple FS shape is very different from those reported in the previous studies on superconducting samples \cite{Matsui.anisotropy}, in which the intensity is suppressed at the ``hot spots" like the weakly annealed sample reported in the present work. The change induced by the protect annealing is clear also in the band image plots (Figs.~\ref{fig2}d-f), and corresponding EDCs (Figs.~\ref{fig2}g-i) along the cuts through the node, the ``hot spot", and the anti-node for each sample. At the ``hot spot" of the as-grown and weakly annealed samples, the peak is shifted from $E\mathrm{_F}$ towards higher binding energies and at the anti-node the quasi-particle (QP) peak is split, which can be attributed to AFM correlation. Similar results have also been reported for superconducting samples reduction-annealed by the conventional method \cite{Matsui.anisotropy} as shown in Fig.~\ref{fig2}j, indicating that strong AFM correlation persists even in the superconducting samples. On the other hand, the protect-annealed sample shows that a sharp QP peak disperses towards the Fermi level without splitting in all the cuts, and the AFM pseudogap is totally absent.

\begin{figure}
	\begin{center}
		\includegraphics[width=85mm]{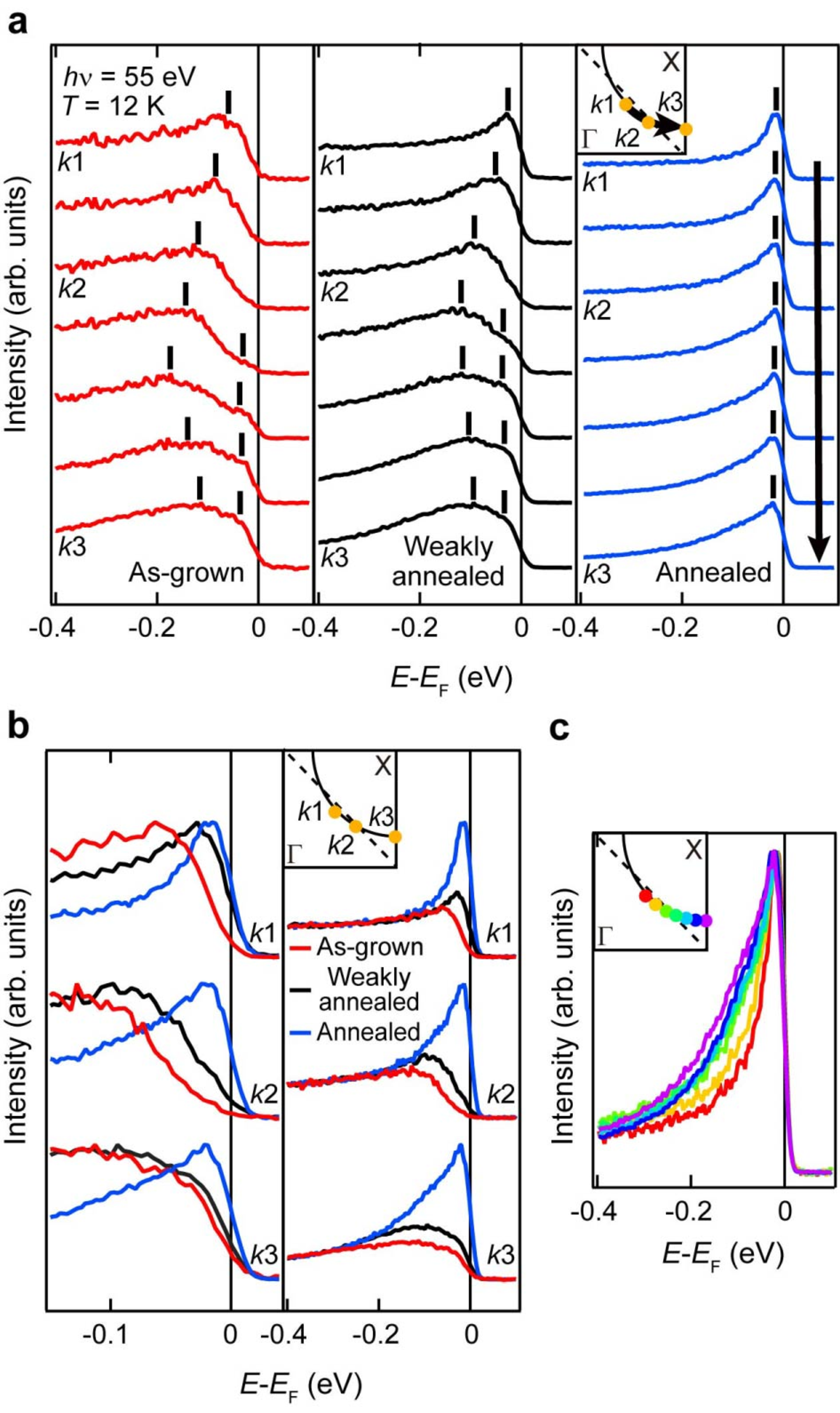}	
	\end{center}
	\caption{{\bf Energy distribution curves (EDCs) on the Fermi surface of PLCCO with and without protect annealing.} {\bf a,} EDCs of as-grown, weakly annealed, and annealed samples, respectively from left to right, along the Fermi surface (for $k_\mathrm{F}$ positions, see the inset). Peak positions are denoted by vertical bars. {\bf b,}  Evolution of EDC with protect annealing. Left panel shows EDCs normalized to the peak height. By protect annealing, the gap/pseudogap closes on the entire Fermi surface. Right panel shows EDCs normalized to the intensity around -0.4 eV. This plot indicates that the spectral weight of the quasi-particle (QP) peak is dramatically enhanced by annealing. The momentum positions where the EDCs were taken are indicated in the inset. {\bf c,} EDCs on the Fermi surface of the annealed sample plotted without offsets. EDCs are normalized to the peak height after the EDC near ($\pi$,$\pi$) was subtracted as a background. The inset shows the corresponding momentum positions.}
	\label{fig3}
\end{figure}

In Fig.~\ref{fig3}a, EDCs are plotted along the FS for each sample. The as-grown and weakly annealed samples show a pseudogap between the node and the ``hot spot", and band splitting between the ``hot spot" and the anti-node. These features are explained by strong AFM correlation as reported in previous ARPES studies \cite{armitage1,Matsui.anisotropy}. Surprisingly, all of those features are absent in the annealed sample, and a sharp single QP peak is observed on the entire FS, indicating the suppression of AFM correlation. \\

The same EDCs are plotted in Fig.~\ref{fig3}b with different intensity normalizations. The left-hand side of Fig.~\ref{fig3}b, where the EDCs have been normalized to the peak height shows that as-grown sample has a gap on the entire FS, consistent with the transport measurements showing an insulating behavior \cite{Adachi}, and that the gap closes by the annealing, consistent with a previous ARPES measurement reported by Richard {\it et al.} \cite{Richard}. According to the plot in the right panel of Fig.~\ref{fig3}b, where EDCs have been normalized to the intensity around $-0.4$ eV, one can see that the QP peak at the Fermi level on the entire FS is dramatically enhanced by the annealing. This growth of the QP spectral weight suggests that the scattering of the QPs by the residual apical oxygens and other defects is also suppressed by the annealing. 

Suppression of the AFM pseudogap around the ``hot spots" enables us to investigate the low-energy physics on the entire FS. The scattering rate of the QPs $-Z$Im$\Sigma_{\bm{k}}(\epsilon)$, where $\Sigma_{\bm{k}}(\epsilon)$ is the self-energy and $Z$ is the renormalization factor assumed to be constant in the low-energy region considered here, as a function of QP energy $\epsilon$, can be evaluated by multiplying the MDC width $\Delta k$ by $v_\mathrm{F}$ (see Supplementary Information). Figure~\ref{PLCCO_self_cal}a shows thus obtained scattering rate $-Z$Im$\Sigma_{\bm{k}}(\epsilon)$ of the annealed sample with $T_c=27.2$ K (``annealed sample 2") along the three cuts, those crossing the node, the ``hot spot", and the anti-node (Cuts 1, 2, and 3 in Fig.~\ref{fig2}c). The dynamical (i.e., energy-dependent) part of $-Z$Im$\Sigma_{\bm{k}}(\epsilon)$ is also plotted at the bottom of Fig.~\ref{PLCCO_self_cal}a.\par
We consider two possibilities that the QP created by photoemission is scattered by excitations of electron-hole pairs or AFM spin fluctuations. Im$\Sigma_{\bm{k}}(\epsilon)$ for the particle-hole excitation at low temperatures is approximately given by
\begin{align}
\mathrm{Im} \Sigma_{\bm{k}}(\epsilon = E_{\bm{k}} < 0) \propto \sum_{\bm{q},\omega} \mathrm{Im}\chi(\bm{q},\omega) \mathrm{Im}\frac{f(E_{\bm{k}+\bm{q}})}{E_{\bm{k}+\bm{q}}-E_{\bm{k}}-\omega +i\delta},
\label{ImSigma}
\end{align}
where $\chi(\bm{q},\omega)=\sum_{\bm{k'}} \frac{f(E_{\bm{k'}+\bm{q}})-f(E_{\bm{k'}})}{E_{\bm{k'}+\bm{q}}-E_{\bm{k'}}-\omega +i\delta}$ is the Lindhard function \cite{Brinkman,Markiewicz}. Im$\Sigma_{\bm{k}}(\epsilon)$ due to AFM spin fluctuations with finite correlation length $\xi$ and finite spin fluctuation energy $\omega _{\mathrm{SF}}$ is given by the equation (\ref{ImSigma}) with
\begin{align}
\mathrm{Im} \chi(\bm{q},\omega) \propto \frac{\omega}{\{ 1+ (\bm{q} - \bm{Q}_{\mathrm{AFM}})^2 \xi ^2 \}^2 + (\omega / \omega _{\mathrm{SF}})^2},
\label{omega_q}
\end{align}
where $\bm{Q}_{\mathrm{AFM}} \equiv (\pi, \pi)$ \cite{Millis,Zha}. Using equation (\ref{omega_q}) with $\omega _{\mathrm{SF}} = 6$ meV deduced from the inelastic neutron scattering measurement of Pr$_{1-x}$LaCe$_{x}$CuO$_4$ \cite{Fujita.PLCCO}, and the experimentally obtained band structure $\epsilon=E_{\bm{k}}$ fitted to the tight-binding model \cite{Ikeda.t'} (see Supplementary Information), $\mathrm{Im} \Sigma_{\bm{k}}(\epsilon = E_{\bm{k}})$ was calculated along the three cuts for different $\xi$ values (for the detail of the calculation, see Supplementary Information). The calculated $-\mathrm{Im} \Sigma_{\bm{k}}(\epsilon = E_{\bm{k}})$ for AFM spin fluctuations and particle-hole excitations are shown in Figs.~\ref{PLCCO_self_cal}b and c, respectively. AFM spin fluctuations with $\xi \gtrsim 4a$ ($a$: in-plane lattice constant) yield strong scattering around the ``hot spot" in the low energy region because low energy AFM spin fluctuations scatter the QPs near one ``hot spot" to another. However, when the correlation length is decreased to $\xi = 2a$, the scattering at the ``hot spot" is no longer clear as is the case for the scattering by particle-hole excitations. The dynamical part of the QP scattering $-(Z$Im$\Sigma_{\bm{k}}(\epsilon)-Z$Im$\Sigma_{\bm{k}}(\epsilon = 0)$), i.e., the inelastic scattering of QPs, is almost the same among the three cuts in the experiment as shown in Fig.~\ref{PLCCO_self_cal}a, suggesting that the AFM correlation length in the protect-annealed sample is $\xi \lesssim 2a$. A calculation based on an AFM-phase fluctuation model \cite{Park_phase_fluctuation} has also shown that if the correlation length is as small as $\xi \sim 2a$, an AFM pseudogap becomes invisible as in the spectra of the protect-annealed samples (Figs.~\ref{fig2} and \ref{fig3}), while an AFM pseudogap opens for $\xi \gtrsim 4a$. The reduction of the magnitude of the fluctuating spin moment would also contribute to the weakening of the AFM pseudogap. If the latter is the case, the AFM correlation length $\xi$ somewhat larger than $2a$ would still be consistent with the ARPES spectra. Thus the absence of AFM correlation signals in the protect-annealed samples indicates that AFM correlation length $\xi$ and/or the magnitude of the (fluctuating) local magnetic moments are dramatically reduced by the protect annealing.

As for the static part of the QP scattering rate $\Gamma_0 = -Z$Im$\Sigma_{\bm{k}}(\epsilon = 0)$, i.e., the elastic scattering rate of QPs, Fig.~\ref{PLCCO_self_cal}a indicates its enhancement near the anti-node. Note that the elastic scattering is caused by static disorder and should be added to the dynamical scattering, which vanishes at $\epsilon = 0$. In fact, Fig.~\ref{PLCCO_self_cal}d shows that $\Gamma_0$ obtained by fitting $-Z$Im$\Sigma_{\bm{k}}(\epsilon)$ to a power law function $\Gamma_0 + A \epsilon ^\alpha$ (see Supplementary Information) increases as one approaches the anti-node. Consistent with these data, the EDC width is also broader around the anti-node as one can see from Fig.~\ref{fig3}c, suggesting stronger QP scattering in the anti-nodal region. This tendency has been widely observed in the hole-doped cuprates \cite{Valla.prl,Kaminski,Plate,Yoshida}, suggesting common QP scattering mechanisms both in the hole- and electron-doped cuprates. As for the hole-doped cuprates, coupling with AFM fluctuations peaked at $(\pi,\pi)$ \cite{Shen_Schrieffer} or scattering between van Hove singularities (e.g., between $(\pi,0)$ and $(0,\pi)$) \cite{Furukawa_Rice} have been proposed as a possible origin. In the case of e-HTSCs, however, the $(\pi,\pi)$ scattering mechanism is less effective because the wave vector connecting two anti-nodal parts of the FS are strongly deviated from $(\pi,\pi)$ because of the smaller radius of the FS compared to those of hole-doped cuprates (Fig.~\ref{PLCCO_self_cal}e). The van Hove singularity scenario is also difficult for e-HTSCs because the singularities lie well ($\sim$ 400 meV) below $E_\mathrm{F}$ as opposed to $\sim$ 100 meV in the hole-doped cuprates. Alternatively, weak nesting between two anti-nodal parts of the FS around $(\pi,0)$ could enhance elastic scattering of the QPs. If such scattering is strong, incipient charge instability may arise from this FS nesting (Fig.~\ref{PLCCO_self_cal}e). Recently, charge ordering with $\bm{q} \sim (0.25\pi, 0)$ was indeed found both in hole- and electron-doped cuprates \cite{Comin,daSilva}. As for the electron-doped cuprates, the $q$ vector is reported to connect either two anti-nodal points or ``hot spots", and hence it is possible that QPs are scattered between two anti-nodal regions connected by $\bm{q} \sim (0.25\pi, 0)$ and the same scattering causes charge instability.

It is interesting to discuss the possible relevance of the present result to the superconductivity with much lower Ce concentration or even without Ce doping reported for thin films and powdered samples of e-HTSCs \cite{Tsukada.Nondoped, Matsumoto.NCCO, Asai.powder, Takamatsu.powder}. In those samples, superconductivity with $T_c$ as high as the present protect-annealed samples is achieved rather independently of the Ce concentration. Recently, it has been proposed using the local-density approximation combined with dynamical mean-field theory that the AFM long-ranged order is necessary to open a charge-transfer gap in the parent compound of e-HTSCs while Coulomb repulsion without AFM order is sufficient to open the gap in the hole-doped cuprates \cite{Weber.Nature,Weber.PRB}. In addition, it has been shown that, when protect-annealed, even extremely underdoped bulk single-crystalline PLCCO ($x = 0.05$) becomes superconducting with $T_c$'s comparable to the present annealed samples (T.A., A.T., M. Ohgi, and Y.K. unpublished., shown in Fig.~\ref{fig4}).

The doped electron concentrations of the as-grown and weakly annealed samples estimated from the FS area, $n_\mathrm{FS}$'s, were 0.131 and 0.130, both of which were not far from the nominal Ce concentration $x = 0.10$ (see Supplementary Information). On the other hand, the $n_\mathrm{FS}$'s of the protect-annealed samples fell in the range from $n_\mathrm{FS}$ = 0.12 to 0.185, some of which are significantly larger than that expected from the nominal Ce concentration $x = 0.10$ (see Supplementary Information). In Fig.~\ref{fig4}, the $T_c$'s of the protect-annealed samples are plotted against the electron concentration. For comparison, the $T_c$'s of PLCCO and Pr$_{1-x}$LaCe$_x$CuO$_4$ single crystals annealed by the conventional method \cite{Sun,Fujita.PLCCO,Shan} are also plotted with respect to the nominal Ce concentration $x$. In the previous studies, the $T_c$ rapidly decreases with increasing Ce concentration above $x \sim 0.11$. On the other hand, the present samples maintain high $T_c$'s compared to all the previous samples up to the highest $n_\mathrm{FS}$ of 0.185 as shown in Fig.~\ref{fig4}. This can be understood under the assumptions that Ce doping causes structural disorder and that high $T_c$ can be maintained if more electrons can be doped without increasing Ce concentration (beyond $x = 0.10$).

Finally, we discuss possible origins of the excess electron doping determined by ARPES compared to the doped Ce content. Since the protect annealing reduces the oxygen content by $\sim 0.04$ (see Methods), the additional electrons should have been introduced by oxygen removal either from the CuO$_2$ planes or the (Pr,La,Ce)O layers. Although one cannot identify the position of the removed oxygen atoms at present, considering the relatively high $T_c$ of the protect-annealed sample, one can conclude that oxygen removal takes place at atomic sites which induce less disorder than Ce substitution.

In conclusion, we have performed ARPES measurements on protect-annealed PLCCO single crystals with Ce doping of $x = 0.10$ with varying annealing conditions. Sufficiently annealed samples showed a $T_c$ as high as 27.0 K and did not show any signature of AFM fluctuations or the AFM pseudogap, which has been observed in all the other e-HTSCs so far. While the scattering of QPs near $E_\mathrm{F}$ by AFM correlation was not observed at the ``hot spot" in the annealed samples, stronger scattering was observed in the anti-nodal region than in the nodal region, similar to the hole-doped cuprates. This suggests the existence of common scattering mechanisms both in the hole- and electron-doped cuprates although the $(\pi, \pi)$ scattering and the van Hove singularity mechanisms proposed for the hole-doped cuprates do not seem important for the electron-doped cuprates. The protect-annealed samples studied here showed almost the same $T_c$'s, whereas the actual electron concentration estimated from the FS area varied over a wide range. Thus the intrinsic electronic structure revealed by the present ARPES study will be of great importance to elucidate the mechanism of the high-temperature superconductivity.

.

\section*{Methods}
Single crystals of PLCCO with $x = 0.10$ were synthesized by the traveling-solvent floating-zone method. The composition of the as-grown crystals was determined to be Pr$_{1.17}$La$_{0.73}$Ce$_{0.10}$Cu$_{1.00}$O$_{4+\delta}$ by the inductively coupled plasma (ICP) method by assuming that the content of Pr, La, Ce, and Cu amounts to be 3 atoms/f.u. Three kinds of samples were prepared; as-grown, weakly annealed, and annealed samples, among which only the annealed ones showed superconductivity with the  $T_c$ of 27.0 K (``annealed sample 1"). By protect annealing, the oxygen content was reduced by 0.04. We also prepared two additional annealed samples with $T_c$ = 27.2 K and 26.2 K (``annealed sample 2" and ``annealed sample 3" samples, respectively). The ``weakly annealed samples" were annealed at 650 Ž for 24 hours, and the ``annealed samples" at 800 Ž for 24 hours. A study using X-ray and neutron scattering observed Cu defects in as-grown Pr$_{1-x}$LaCe$_x$CuO$_4$ ($x = 0.12$), and an impurity phase of rare-earth oxides in annealed ones, concluding that the annealing repaired the Cu defects with creating new impurity phase, instead of removing apical oxygen \cite{Kang_anneal}. On the other hand in the present samples, Cu deficiency was not detected in as-grown ones by the above mentioned ICP analysis. Furthermore, an impurity phase of rare-earth oxides in annealed samples was not detected by scanning electron microscopy (SEM) \cite{Adachi}. ARPES experiment was performed at beamline 28A of Photon Factory and beamline 9A of Hiroshima Synchrotron Radiation Center (HiSOR). The total energy resolution was set at 28 and 8 meV, respectively. The samples were cleaved {\it in situ}. The measurements were performed under the pressure better than $2 \times 10^{-10} \ \mathrm{Torr}$ and $5 \times 10^{-11} \ \mathrm{Torr}$, respectively. Temperature during the measurement was set to 12 K at Photon Factory, and 9 K at HiSOR.

\section*{Acknowledgements}
Useful discussion with M. Ogata, C. M. Varma, and T. Saha-Dasgupta is gratefully acknowledged. ARPES experiments were performed at KEK-PF (Proposal Nos. 2012G075, 2014G177 and 2012S2-001) and HiSOR (Proposal Nos. 12-A-20 and 14-A-13). This work was supported by Grants-in-Aid for Scientific Research on Innovative Areas ``Frontier of Materials, Life and Particle Science Explored by Ultra Slow Muon Microscope" and ``Materials Design through Computics" from MEXT, Japan.

\section*{Author contributions}
M.H., T.Y., H.S., L.C.C.A., K. Okazaki., D.O., K.S., M.T., and T.M. performed ARPES measurements with the assistance of H.K., K. Ono, H.A., and M.A. M.H. analyzed the data and performed the calculations. Y.M., A.T., T.A., and Y.K. synthesized and characterized single crystals. M.H. and A.F. wrote the manuscript with suggestions by T.A., Y.K., T.Y., T.M., and all other coauthors. A.F. was responsible for overall project direction and planning.

\section*{Competing financial interests}
The authors declare no competing financial interests.

\newpage.

\begin{figure}[!h]
	\begin{center}
		\includegraphics[width=150mm]{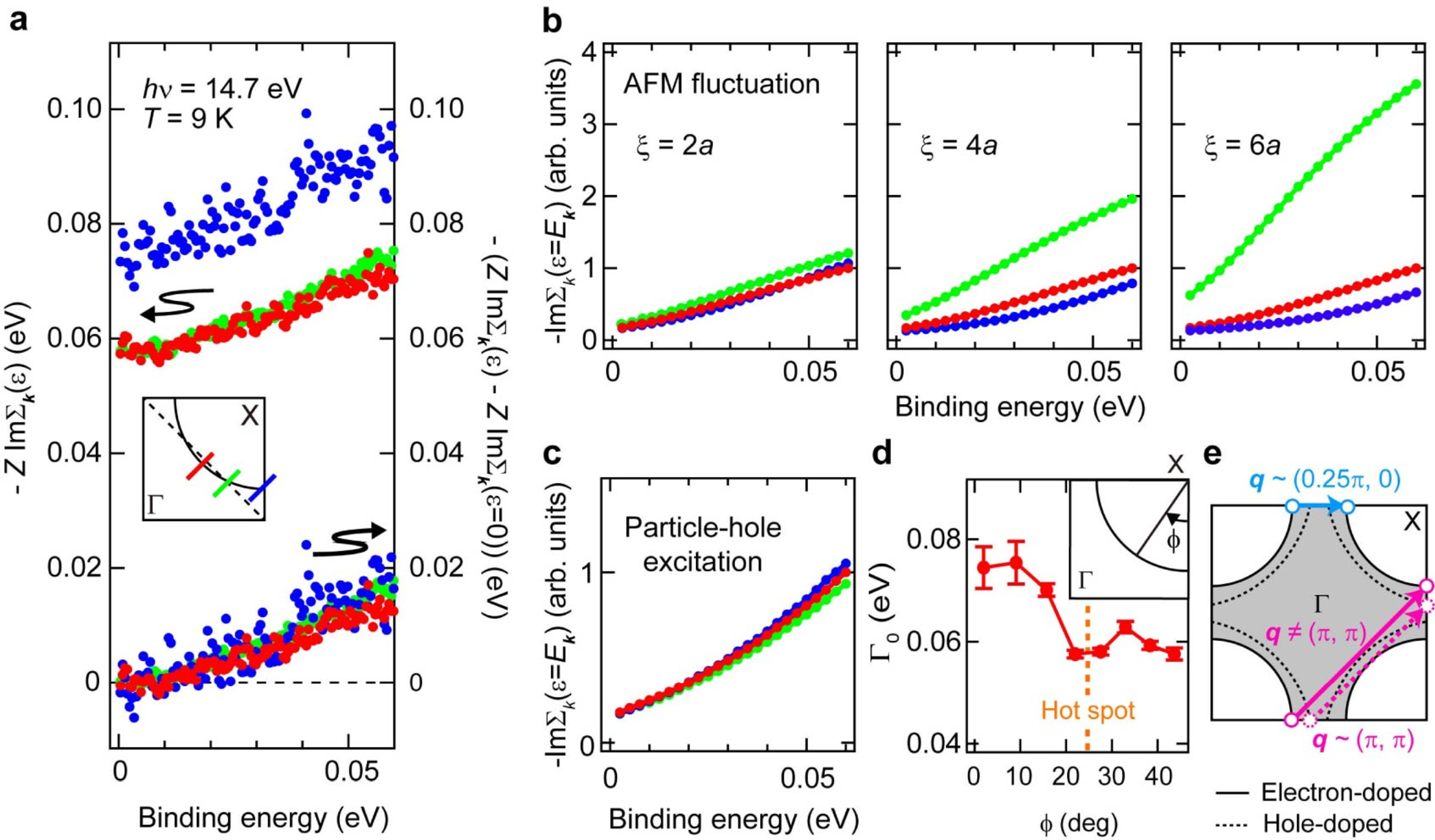}	
	\end{center}
	\caption{{\bf Scattering rate of quasi-particles (QPs) near the Fermi level in protect-annealed samples.} {\bf a,} Experimentally obtained scattering rate $-Z$Im$\Sigma_{\bm{k}}(\epsilon)$ of QPs for the protect-annealed sample with $T_c=27.2$ K (annealed sample 2) along the cuts indicated in the inset. The dynamical part of the scattering rate $-(Z$Im$\Sigma_{\bm{k}}(\epsilon)-Z$Im$\Sigma_{\bm{k}}(\epsilon = 0)$) is also plotted at the bottom. {\bf b,} Simulation of the dynamical scattering rate $-$Im$\Sigma_{\bm{k}}(\epsilon = E_{\bm{k}})$ along the cuts indicated in the inset in {\bf a} for AFM fluctuations with the correlation length of $\xi = 2a$ (left), $4a$ (middle), and $6a$ (right). Calculated $-$Im$\Sigma_{\bm{k}}(\epsilon = E_{\bm{k}})$ has been normalized to the value at the binding energy of 0.06 eV in the nodal cut. {\bf c,} The same plot as {\bf b} for particle-hole excitations. {\bf d,} Elastic scattering rate $\Gamma_0$ plotted against the Fermi surface angle $\phi$ defined in the inset. The position of the ``hot spot" is indicated by a dashed vertical line. {\bf e,} Schematic drawing of elastic scattering of QPs near the anti-node. Solid and dashed curves represent the FSs of electron- and hole-doped cuprates, respectively. Solid and dashed pink arrows correspond to the wave vectors which connects two anti-nodal parts on the FS of the electron- and hole-doped cuprates, respectively. A blue arrow is ``nesting vector" connecting two anti-nodal regions that may lead to charge instabilities in e-HTSCs.}
	\label{PLCCO_self_cal}
\end{figure}

\begin{figure}[!h]
	\begin{center}
		\includegraphics[width=100mm]{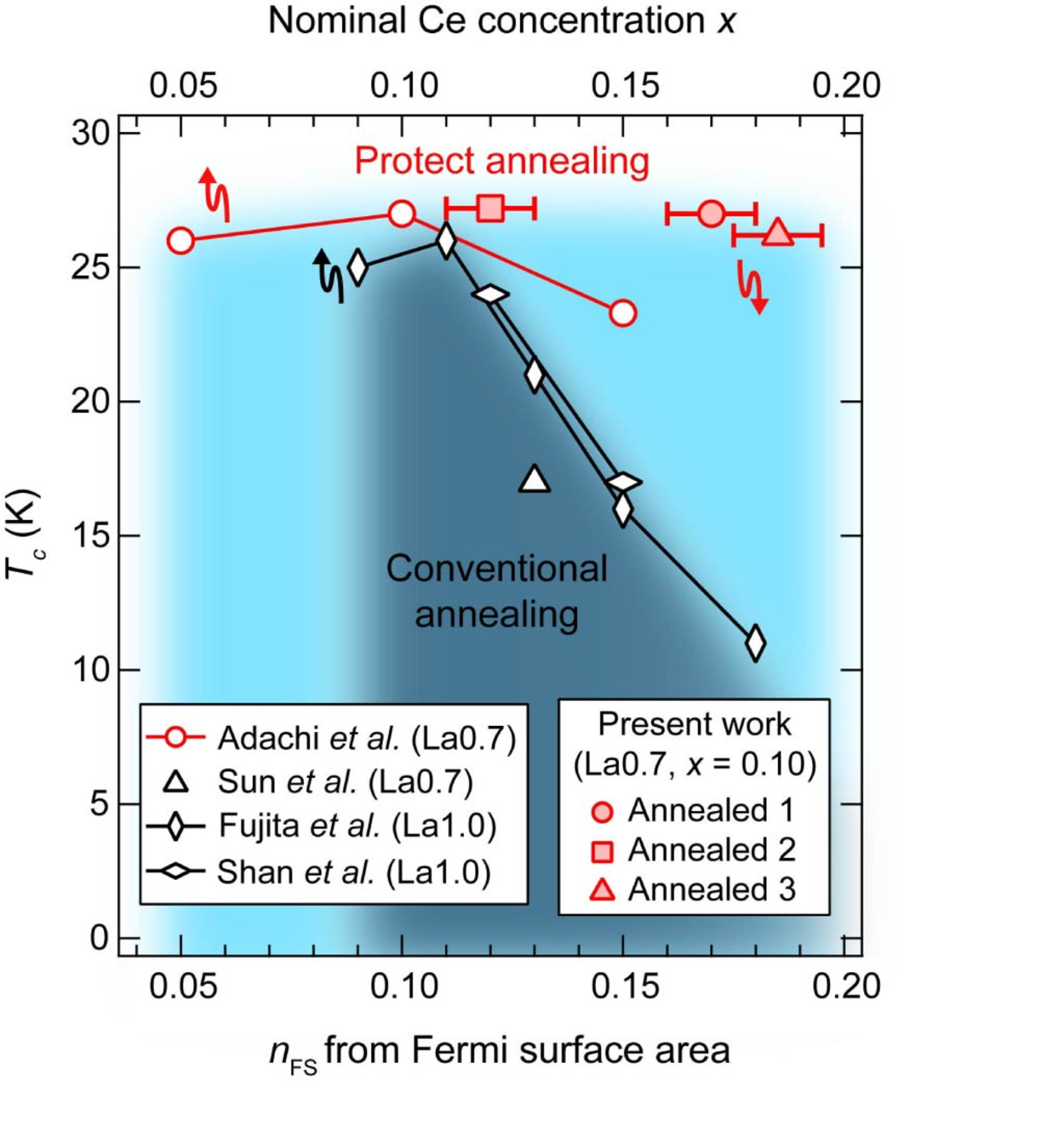}	
	\end{center}
	\caption{{\bf $T_c$ versus Fermi surface area of PLCCO.} The $T_c$'s of three protect-annealed samples plotted against the doped electron concentration, $n_\mathrm{FS}$, estimated from the area of the Fermi surface. For comparison, the $T_c$'s of PLCCO and Pr$_{1-x}$LaCe$_x$CuO$_4$ single crystals annealed by the conventional method \cite{Sun,Fujita.PLCCO,Shan}, and those of the protect-annealed PLCCO single crystals (T.A., A.T., M. Ohgi, and Y.K. unpublished.) are also plotted against the Ce concentration $x$. The data for PLCCO and Pr$_{1-x}$LaCe$_x$CuO$_4$ are denoted by La0.7 and La1.0, respectively.}
	\label{fig4}
\end{figure}

\clearpage

\renewcommand{\thefigure}{S\arabic{figure}}
\renewcommand{\thesection}{\arabic{section}}
\renewcommand{\theequation}{S\arabic{equation}}
\let\MakeTextUppercase\relax
\setcounter{figure}{0}
\setcounter{equation}{0}

\title{Supplementary Information \vskip1cm Suppression of the antiferromagnetic pseudogap in the electron-doped high-temperature superconductor by \\ ``protect annealing" \vspace{1cm}}


\begingroup
\let\clearpage\relax
\let\vfil\relax
\maketitle
\endgroup

\noindent
{\it \footnotesize $^1$Department of Physics, University of Tokyo, Bunkyo-ku, Tokyo 113-0033, Japan}  \\
{\it \footnotesize $^2$Department of Engineering and Applied Sciences, Sophia University, Tokyo 102-8554, Japan} \\
{\it \footnotesize $^3$Department of Applied Physics, Tohoku University, Sendai 980-8579, Japan} \\
{\it \footnotesize $^4$KEK, Photon Factory, Tsukuba 305-0801, Japan} \\
{\it \footnotesize $^5$Hiroshima Synchrotron Radiation Center, Hiroshima University, Higashi-Hiroshima 739-0046, Japan} \\
{\it \footnotesize $^6$Graduate School of Science, Hiroshima University, Higashi-Hiroshima 739-8526, Japan} \\
{\it \footnotesize $^7$Graduate School of Frontier Sciences, University of Tokyo, Kashiwa 277-0882, Japan} \\

\vfill

\noindent
$^*$e-mail: horio@wyvern.phys.s.u-tokyo.ac.jp \\
{\footnotesize $^{\dag}$Present address: Graduate School of Engineering, Osaka Prefecture University, Sakai 599-8531, Japan \\}

\newpage

\begingroup
\section{Experimental estimates and simulation of the scattering rate of the quasi-particles}
\endgroup
The scattering rate of the quasi-particles $-Z$Im$\Sigma_{\bm{k}}(\epsilon)$, where $\Sigma_{\bm{k}}(\epsilon)$ is the self-energy and $Z$ is the renormalization factor assumed to be constant in the low-energy region considered here, as a function of QP energy $\epsilon$, was evaluated from the ARPES spectra by multiplying the MDC width $\Delta k$ by $v_\mathrm{F}$. $\Delta k$ was estimated by fitting the MDC at each energy to a Lorentzian, and $v_\mathrm{F}$ was determined by fitting the band dispersion from $E_\mathrm{F}-35$ meV to $E_\mathrm{F}-5$ meV, below which a kink was observed \cite{Park_kink_sup}. We regarded $v_\mathrm{F}$ as constant within this energy range. The $\Delta k$ and $v_\mathrm{F}$ values have been corrected for the angle between the cut direction and the FS normal when they are not parallel to each other. Thus obtained $v_\mathrm{F}$ was 2.1 eV \AA, 2.4 eV \AA \ and 1.9 eV \AA \ at the node, the ``hot spot", and the anti-node, respectively.

To determine the elastic scattering rate $\Gamma_0$, $-Z$Im$\Sigma_{\bm{k}}(\epsilon)$ was fitted to the power law function $\Gamma_0 + A \epsilon ^\alpha$ in the energy range from $E_\mathrm{F}$ to 35 meV below it. The error bar was determined by the 3$\sigma$ of the fitting.

In calculating $-$Im$\Sigma_{\bm{k}}(\epsilon = E_{\bm{k}})$ using equation (1) and (2) of the main text, two-dimensional $k$ space and $q$ space were covered by a mesh of $400 \times 400$, $\omega$ from 0 eV to 0.1 eV was divided at 5 meV intervals, and $\delta$ was set to 0.01 eV. Temperature was set to 9 K, the same condition as the experiment. Calculated $-$Im$\Sigma_{\bm{k}}(\epsilon = E_{\bm{k}})$ has been normalized to the value at the binding energy of 0.06 eV in the nodal cut.\\

\section{Tight-binding fit of the Fermi surface and band dispersions}
\begin{figure}
	\begin{center}
		\includegraphics[width=150mm]{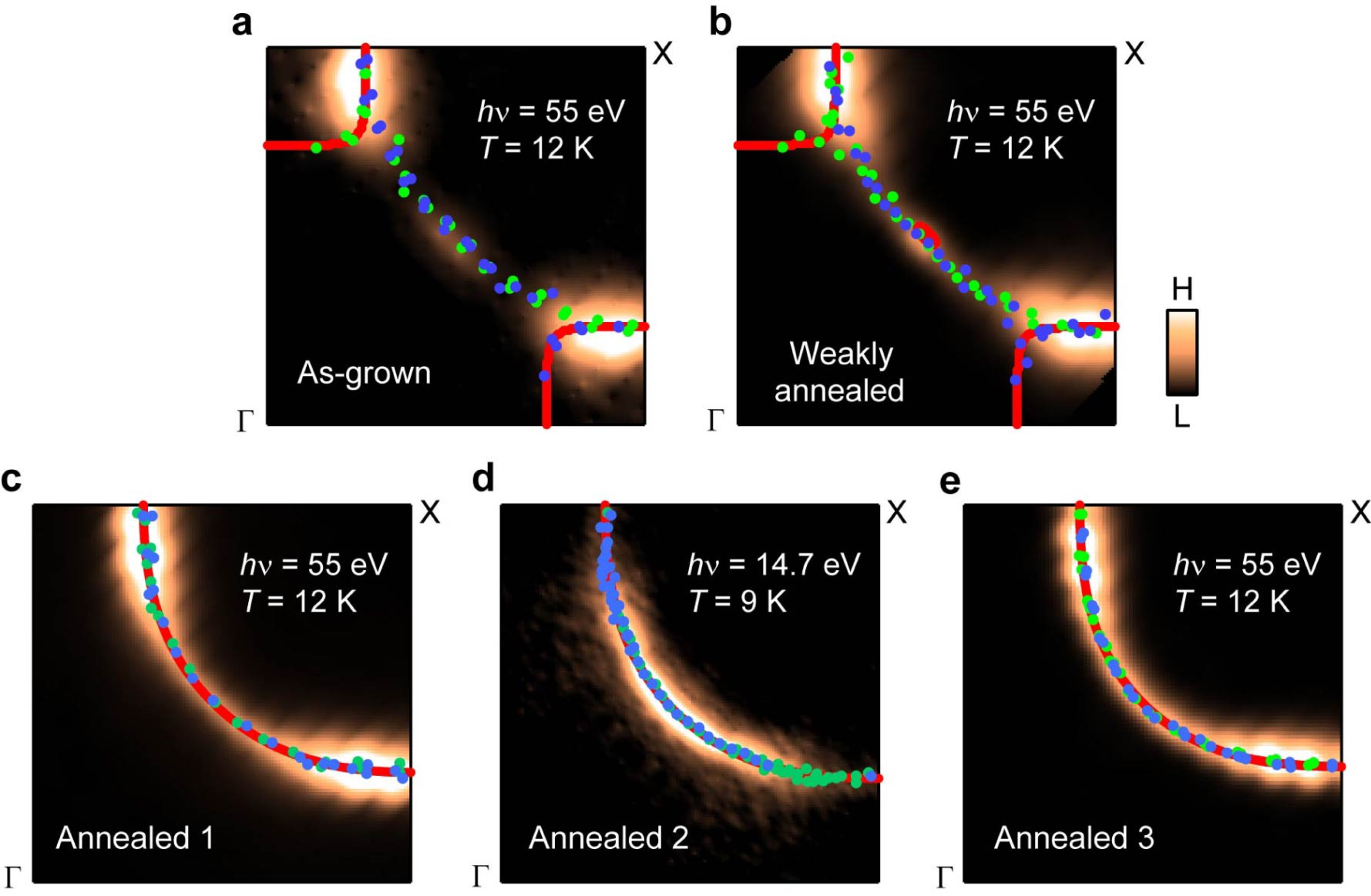}
	\end{center}
	\caption{{\bf Tight-binding fit of the Fermi surface.} {\bf a, b,} Symmetrized Fermi surface (FS) mappings fitted to the tight-binding model displayed by red curves for the as-grown and weakly annealed Pr$_{1.3-x}$La$_{0.7}$Ce$_x$CuO$_4$ (PLCCO, $x=0.10$) samples, respectively. Blue points are the peak positions of the momentum distribution curves (MDCs) at $E_\mathrm{F}$ which are obtained in the displayed momentum region or from outside of the displayed region by symmetry operations considering the four-fold symmetry. Green points have been obtained by symmetrizing blue ones with respect to the $\Gamma$-X line. {\bf c-e,} The same plots as {\bf a} and {\bf b} for annealed sample 1, 2, and 3, respectively.}
	\label{TB_sup1}
\end{figure}

\begin{figure}
	\begin{center}
		\includegraphics[width=130mm]{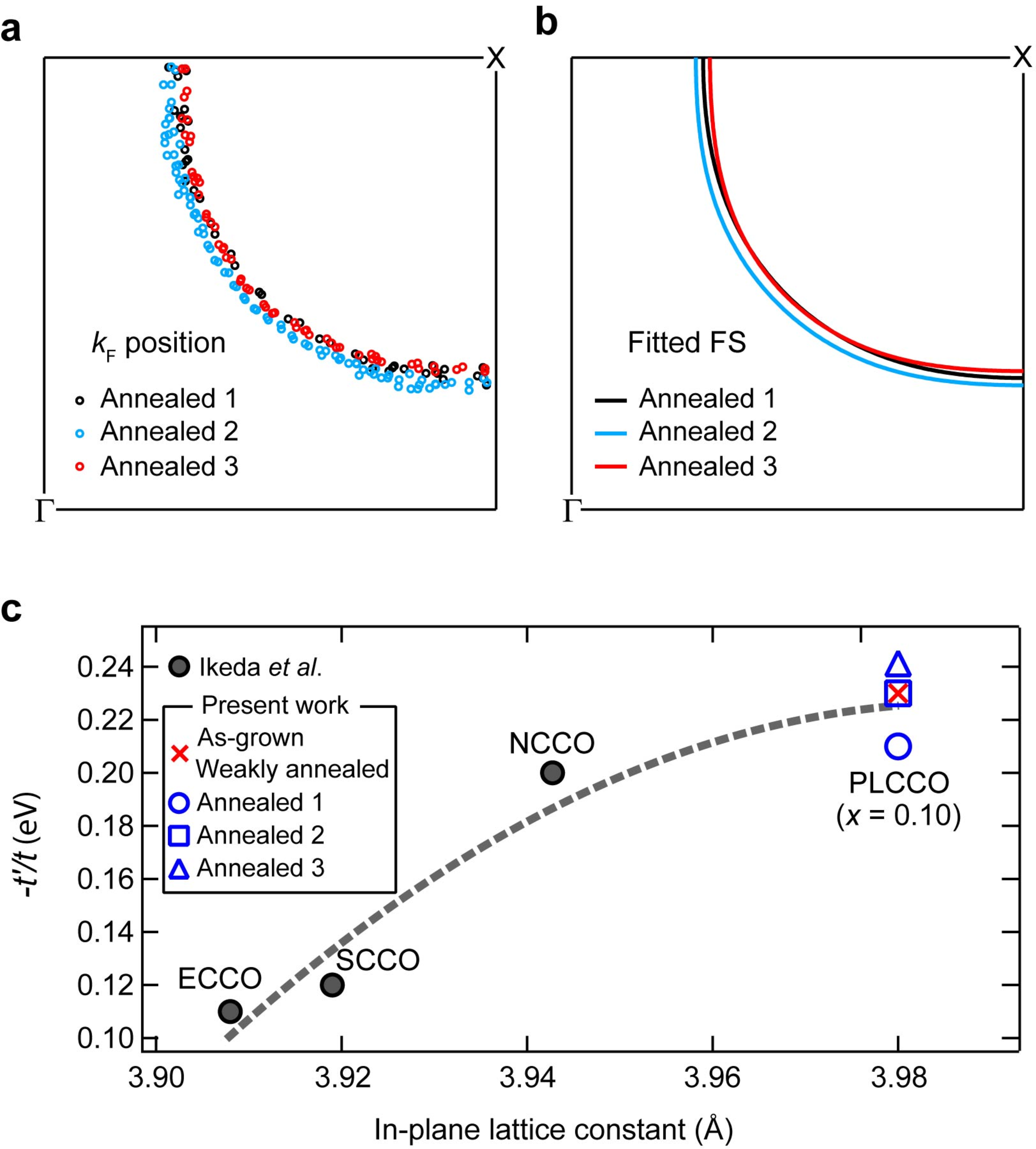}
	\end{center}
	\caption{{\bf Shape of the Fermi surface.} {\bf a,} $k_\mathrm{F}$ positions extracted from Figs.~\ref{TB_sup1}{\bf c-e} for three annealed samples. {\bf b,} FSs obtained by fitting $k_\mathrm{F}$ positions plotted in {\bf a} to the tight-binding model. {\bf c,} Relationship between $-t'/t$ and the in-plane lattice constant. The data for $\mathrm{Nd_{1.85}Ce_{0.15}CuO_4}$(NCCO), $\mathrm{Sm_{1.85}Ce_{0.15}CuO_4}$(SCCO), and $\mathrm{Eu_{1.85}Ce_{0.15}CuO_4}$(ECCO) are taken from Ikeda {\it et al.} \cite{Ikeda.t'_sup}.}
	\label{TB_sup2}
\end{figure}

The FSs and band dispersions of the as-grown and weakly annealed samples were fitted to the tight-binding model of the square lattice consisting of the Cu $d_{x^2-y^2}$ orbitals with the $\sqrt{2} \times \sqrt{2}$ AFM order as 
\begin{eqnarray}
\epsilon &-& \mu = \epsilon _0 \pm \sqrt{\Delta ^2 + 4t^2 (\mathrm{cos} \ k_xa + \mathrm{cos} \ k_ya)^2} \nonumber \\ 
 &-&4t'\mathrm{cos} \ k_xa \ \mathrm{cos} \ k_ya -2t"(\mathrm{cos} \ 2k_xa + \mathrm{cos} \ 2k_ya),
\end{eqnarray}
where $t$, $t'$, and $t"$ are the nearest-neighbor, next-nearest-neighbor, and third-nearest-neighbor transfer integrals, $\pm \Delta$ denotes the staggered potential of the two sublattices (Figs.~\ref{TB_sup1}a and b). Although this tight-binding model can capture characteristic features of the band structure of e-HTSCs, where AFM correlation is strong, perfect fitting has been difficult because the antiferromagnetism is not a long-ranged one but short-ranged one \cite{Park.short.range_sup} and the AFM gap $\Delta$ is generally $\bm{k}$-dependent \cite{Matsui.anisotropy_sup}, probably due to complicated electron correlation effect that is not considered in the simple AFM tight-binding model. In fact, a variational Monte-Carlo calculation \cite{Chou.Lee_sup} has shown that the ``AFM gap" takes the largest value at ($\pi$/2, $\pi$/2), and the smallest value of almost zero at ($\pi$, 0) and (0, $\pi$), and hence the dispersion around the band bottom at ($\pi$, 0) and (0, $\pi$) can be rather well fitted to the tight-binding model with $\Delta = 0$ \cite{Ikeda.t'_sup}. Therefore, for the present as-grown and weakly annealed samples, we determined the value of the parameter $t'$ from the energy position of the band bottom at ($\pi$, 0) and (0, $\pi$) using the tight-binding model with $\Delta = 0$. On the other hand, the FS of the annealed samples can be well fitted to the tight-binding model without $\Delta$, that is, without any signature of AFM correlation, as shown in Figs.~\ref{TB_sup1}c-e. The suppression of the intensity at the ``hot spot" could not be detected in all the three annealed samples, and the FSs could be well fitted to the $\Delta = 0$ tight-binding model. The size of the FS varied among different annealed samples as one can see from Figs.~\ref{TB_sup2}a and b. The doped electron concentrations estimated from the area of the fitted FS, $n_\mathrm{FS}$'s, were 0.175, 0.120, and 0.185 for annealed sample 1, 2, and 3, respectively. The parameter $-t'/t$, which represents the curvature of the FS, are plotted in Fig.~\ref{TB_sup2}c together with those deduced from other e-HTSCs \cite{Ikeda.t'_sup}. Here, $-t"/t'$ has been fixed at 0.5. The $t'/t$ value of the present PLCCO samples follows the tendency against the in-plane lattice constant reported by Ikeda {\it et al.} \cite{Ikeda.t'_sup} regardless of the extent of the annealing.

\newpage

\newpage

\end{document}